# AGuIX nanoparticle-nanobody bioconjugates to target immune checkpoint receptors

Léna Carmès [a,b*] , Guillaume Bort [a*,c,d] , François Lux [a,e†], Léa Seban [f] , Paul Rocchi [a,b], Zeinaf Muradova [f] , Agnes Hagège [g] , Laurence Heinrich-Balard [h] , Frédéric Delolme [i] , Virginie Gueguen-chaignon [j] , Charles Truillet [k] , Stephanie Crowley [l], Elisa Belllo [l] , Tristan Doussineau [b] , Michael Dougan [l,m], Olivier Tillement [a] , Jonathan D. Schoenfeld [f] , Needa Brown [f,m†] and Ross Berbeco [f†]

a. Institut Lumière-Matière, UMR 5306, Université Lyon1-CNRS, Université de Lyon, Villeurbanne Cedex 69100, France
b. NH TherAguix SA, Meylan 38240, France
c. Institut Universitaire de France (IUF), Paris, France
d. Institut Curie, PSL Research University, CNRS, INSERM, UMR9187, U1196, F91405, Orsay, France
e. Université Paris Sud, Université Paris-Saclay, CNRS, UMR9187, U1196, F-91405 Orsay, France
f. Department of Radiation Oncology, Brigham and Women's Hospital, Dana-Farber Cancer Institute, and Harvard Medical School, Boston 02115, USA.
g. Université de Lyon, CNRS, Université Claude Bernard Lyon 1, Institut des Sciences Analytiques, UMR 5280, 69100, Villeurbanne, France
h. Université Lyon 1, CNRS, MATEIS, UMR5510, Univ Lyon, Université Claude Bernard Lyon 1, Villeurbanne 69100, France
i. Univ Lyon, Université Claude Bernard Lyon 1, ENS de Lyon, CNRS UAR3444, Inserm US8, SFR Biosciences, F-69007 Lyon, France
j. Université Paris-Saclay, CEA, CNRS, Inserm, BioMaps, SHFJ, Orsay 91400, France
k. Whitehead Institute for Biomedical Research, Cambridge, MA, 02142, USA
l. Division of Gastroenterology, Massachusetts General Hospital, Boston, MA, 02114, USA
m. Department of Physics, Northeastern University, Boston 02115, USA.
* These authors contributed equally to this work.
† Co-corresponding authors. Ross Berbeco: ross_berbeco@dfci.harvard.edu; Needa Brown: ne.brown@northeastern.edu; François Lux: francois.lux@univ-lyon1.fr

This article presents bioconjugates combining nanoparticles (AGuIX) with nanobodies (VHH) targeting Programmed Death Ligand 1 (PD-L1, A12 VHH) and Cluster of Differentiation 47 (CD47, A4 VHH) for active tumor targeting. AGuIX nanoparticles offer theranostic capabilities and an efficient biodistribution/pharmacokinetic profile (BD/PK), while VHH's reduced size (15 kDa) allows efficient tumor penetration. Site-selective sortagging and click chemistry were compared for bioconjugation. While both methods yielded bioconjugates with similar functionality, click chemistry demonstrated higher yield and could be used for the conjugation of various VHH. The specific targeting of AGuIX@VHH has been demonstrated in both in vitro and ex vivo settings, paving the way for combined targeted immunotherapies, radiotherapy, and cancer imaging.

## 1. Introduction

In recent decades, inorganic nanoparticles (NPs) have demonstrated significant promise for oncologic indications. One noteworthy example is the AGuIX (Activation and Guidance of Irradiation X) NPs. AGuIX are ultra-small NPs composed of gadolinium (Gd) chelates on a polysiloxane core, displaying a mean hydrodynamic diameter of 4 ± 2 nm. The corresponding mean molecular weight (MW) has been estimated at around 20 kDa. These NPs are currently being investigated in several clinical trials targeting different types of cancer, including brain metastasis (Phase II: NCT03818386 and NCT04899908), cervical cancer (Phase I: NCT03308604), glioblastoma (Phase I/II: NCT04881032), and pancreatic and lung cancers (Phase I/II: NCT04789486).[1,2] AGuIX NPs possess theranostic properties, providing dual functionality for imaging and radiotherapy. This is achieved through the presence of paramagnetic gadolinium ions ($Gd^{3+}$) embedded within the NPs. These gadolinium ions contribute to positive contrast effects in T1 magnetic resonance imaging (MRI) and enhance the effectiveness of radiotherapy due to their high atomic number (Z=64). [3,4] The small size of AGuIX NP enables preferential passive uptake in tumors while facilitating rapid elimination through the kidneys(in humans t1/2 = 1.29 ± 0.27 h), thus minimizing the risk of toxicity. Additionally, AGuIX NPs demonstrate tumor uptake comparable to that of other organic or inorganic NPs, achieving approximately 1.19 ± 0.87% ID/g through passive targeting after 24 hours in large variety of animal model. [5,6] Nevertheless, to further enhance the efficacy of AGuIX NP, improving specificity and tumor retention time via active tumor targeting is an important next step for AGuIX translation and future generations.

Preclinical studies employing bioconjugation of AGuIX with antibodies and peptides have showcased the effectiveness of active targeting, resulting in a minimum 1.43-fold increase in tumor accumulation. [5,7–9] Peptides have gained significant popularity as ligands for targeting cancer cells due to their compact size and ease of production. However, they often exhibit low affinity for the target compared to antibodies.[10] Conversely, antibodies have demonstrated remarkable efficacy as targeting molecules, but their larger size hampers tissue penetration and prevents access to certain areas in tumoral tissues. [11] For these reasons, our work focuses on bioconjugates based on nanobodies, otherwise known as Variable Heavy domain of Heavy chain (VHH). These targeting biomolecules have a molecular weight of approximatively 15 kDa and are roughly ten times smaller than monoclonal antibodies, and therefore offer enhanced efficiency in targeting.[12] Their reduced size enables them to readily penetrate tumor tissues and bind to a greater number of a receptors with enhanced affinity and specificity. Furthermore, VHH are swiftly eliminated from the bloodstream, minimizing the risk of toxic accumulation. They possess a remarkable folding capacity and robust physicochemical properties, endowing them with superior stability and excellent solubility.[12–16] These distinctive features enable VHH to overcome some limitations associated with monoclonal antibody therapies.[17,18] It is worth highlighting that both AGuIX NPs (nanoparticles) and VHHs (single-domain antibodies) exhibit a similar compact size, approximately 15 to 20 kDa, endowing them with similar in vivo properties, particularly in terms of tissue penetration, circulation, and clearance. By bioconjugating them, it is possible to combine these benefits in a single product. Herein, we design bioconjugates that target Programmed Death Ligand 1 (PDL1) and the Cluster of Differentiation 47 (CD47) receptors by using the A12 and A4 VHH respectively.[19–21] A12 exhibits high affinity in the low nanomolar range for human and murine PD-L1, making it suitable for targeting its expression. Similarly, A4 exhibits a binding affinity of approximately 10 pM for murine CD47. [21,20]

The PD-L1 ligand and the innate immune regulator CD47 are relevant immunotherapy targets. PD-L1 can be expressed on tumor cells and binds to PD-1 on the surface of T-cells, triggering immune cell exhaustion and inhibiting anti-tumor immunity. Interfering with this negative immune checkpoint, PD-L1 inhibitors have demonstrated efficacy across many different cancer types. Specifically, atezolizumab, an anti-PD-L1 therapy, was approved for treatment of lung cancer in the U.S. in 2016 after showing a 25.4% overall survival rate after 3 years in the treatment of metastatic non-squamous or squamous non-small cell lung cancer (NSCLC). [22] Four years later, in 2020, duvarlumab, a second anti-PD-L1 antibody, was also approved for lung cancer treatment, demonstrating a 5-year overall survival rate of 42.9%, when administered following definitive chemoradiation therapy. [23] These two antibodies have been approved as a second treatment after one or two doses of chemotherapy. However, despite these promising results, the majority of patients with solid tumors do not respond to PD-L1 therapy, highlighting the need for combination approaches and better patient stratification. [24]

CD47 is an important receptor expressed on tumor infiltrating macrophages that interacts with signal regulatory protein alpha (SIRPα). This regulator is overexpressed in various malignancies and prevents phagocytosis by providing a "don't eat me" signal to immune cells. Targeting and blocking CD47 on

tumors restores this phagocytic response which can then upregulate secondary, adaptive responses. [20,25–27]

Antibodies targeting the CD47 receptor are currently being evaluated in multiple clinical trials. [28–30]

The limited efficacy of existing immunotherapy in treating most solid tumors has highlighted the need to better define the tumor immune microenvironment and develop combination treatment approaches. PD-L1 / CD47 AGuIX combinations could help achieve both aims. Three of the approved immune checkpoint inhibitors (ICI) that target the PD-1/PD-L1 axis require biomarker confirmation of PD-L1, highlighting the need for PD-L1 in vivo diagnostic tools.[31] Similarly, visualization of CD47 expression may help identify tumors amenable to macrophage based therapies. Both PD-L1 and CD47 inhibitors have been demonstrated to synergize with radiotherapy in preclinical models[32–35]; enhancing the effectiveness of radiotherapy with AGuIX could further increase the potency of this combination. Therefore, our primary focus was on developing AGuIX NPs that could effectively target and delineate crucial immunomarkers, specifically PD-L1 and CD47. This research serves as a foundational demonstration of the chemical synthesis approach, aimed at creating a platform that harnesses both the diagnostic and therapeutic potential of AGuIX, in conjunction with the immunotherapeutic capabilities of immunocheckpoint targeting.

To achieve this objective, we grafted AGuIX NPs with two different VHHs for targeted applications using two highly specific techniques: 1) sortagging[36–38] and 2) click chemistry. [39–41] The sortagging is based on an enzyme specific reaction and click on an azide-alkyne reaction. [40,42–45] A12 and A4-modified AGuIX were prepared to compare both conjugation techniques and to assess receptor targeting using various *in vitro* assays. For the selected coupling synthesis, the *ex vivo* approach was studied using A12 nanobody in a highly aggressive murine model of melanoma.

## 2. Materials and methods

2.1. AGuIX NPs

The Gd-chelated polysiloxane NPs (AGuIX) were provided by NH TherAguix (Meylan, France) as a lyophilized powder. Their synthesis has been extensively documented in the scientific literature.[46,47] They contain roughly between 10 and 20 Gd atoms per particle, which can be quantified using ICP-MS (Inductively Coupled PlasmaMass Spectrometry).[3,48] The presence of the gadolinium atoms confers to AGuIX radiation dose amplification and MRI contrast properties. Throughout the paper, each AGuIX NPs concentration is stated in g L$^{-1}$ of AGuIX NP or M of Gd element. Within the polysiloxane structure AGuIX contain primary amines in the (3- Aminopropyl)triethoxysilane (APTES) function. [49,50] Each NP is estimated to possess a similar number of amine functions as Gd chelates. [51,52] These amines will be utilized as the main functional groups for all the biofunctionalizations. The AGuIX NPs used in this study have all been pre-grafted on the free amino function present on polysiloxane matrix with the cyanine 5.5 fluorescent dye (detailed synthesis in ESI).

2.2. VHHs

The A12 and A4 VHHs were synthesized at the Massachusetts General Hospital (USA), as described a previously published protocol. [21] A12 and A4 sequences were sub-cloned into the WK6 *E. coli* periplasmic expression vector pHEN6 to enable Gibson cloning and the inclusion of a C-terminal sortase motif and 6xHis tag. *E. coli* containing the plasmid were grown to mid-log phase at 37°C and VHH expression induced with 1 mM IPTG at 30 °C overnight. Centrifugation (5000 x g, 15 mins, 4 °C) was used to harvest the cells and resuspend them in 25 mL 1X TES buffer. Cells were submitted to osmotic shock in 1:4 0.25x TES buffer overnight at 4 °C. The periplasmic fraction was separated by centrifugation (8000 rpm, 30 mins, 4 °C) and loaded onto Ni-NTA beads (Qiagen) and eluted in 50 mM

Tris, pH 8, 150 mM NaCl, 500 mM imidazole. Eluted protein was loaded onto a Superdex 75 10/300 column. Recombinant VHH purity was assessed by SDS-PAGE and concentrated with an Amicon 10,000 kDa filtration unit (Millipore). VHH were stored at -80 °C.

2.3. Synthesize of AGuIX@VHH by sortagging.

AGuIX-Cy5.5-C(W/T)GGG (600 µM GGG(W/T)C final concentration) and VHH (30 or 40 µM final concentration) were introduced in buffer (50 mM Tris-HCl, pH 7.5, 150 mM NaCl). Ligation was triggered (t0) by the addition of StrA7m (2.5 µM final concentration). The mixture was stirred (orbital) at room temperature for 3 h. After reaction, the crude was negatively selected using Ni-NTA bead (400 µL slurry for 1 mg VHH, spin at 800 G for 1 min) to remove 6xHis-tag starting materials (StrA7m, VHH and released VHH residue after ligation), and filtered (MWCO 50 kDa vivaspin, dilution x100,000). AGuIX@VHH was purified by SEC (size exclusion chromatography, method B detailed in ESI) and stored at -20 °C. AGuIX-Cy5.5-C(W/T)GGG preparations and synthesis optimizations are detailed in ESI.

2.4. Synthesize AGuIX@VHH through click chemistry

A12-azide (50 µM, final concentration in reaction) was added to AGuIX-DBCO at (100 AGuIX mg mL$^{-1}$ final concentration in reaction) in 10 mM Phosphate-Buffered Saline (PBS) solution. The solution was stirred for 24 h at 4 °C. A temporal follow-up of the click reaction is monitored by SEC using method A (detailed in ESI). After reaction, AGuIX@VHH was purified by SEC (method B) to remove sub-products (NHS-terminated, azide-PEG4-NHS and AGuIX-DBCO excess) and stored at -20 °C. AGuIX-Cy5.5-DBCO and A12-azide preparations and synthesis optimizations are detailed in ESI.

2.5. Material characterizations

2.5.1. Size Exclusion Chromatography (SEC) UltraViolet/VisibleFluorescence

The Superdex 75 increase 10/300 GL (Cytiva) was used for SEC, and the process was conducted using an isocratic system of elution in acetate buffer 100 mM (pH 4.7) at a flow rate of 0.8 mL min$^{-1}$ (25 µL injected). The G1311A pump (Agilent), Photodiode Array Detector G1315B (Agilent), and fluorescence detector G1321A (Agilent) were employed during the process to record absorbance at 295 nm and fluorescence at 280 nm excitation and 340 nm emission for the method A and fluorescence at 650 nm excitation and 665 nm emission for the method B. The data were acquired using on ChemStation B.04.03 SP1 and analyzed on GraphPad Prism 8.0.1. For preparative purification, the same column was employed within an AKTA PURE chromatography system (Cytiva) using an isocratic system of elution in phosphate-buffered saline (PBS) 100 mM at a flow rate of 0.8 mL min$^{-1}$ (5 mL injected).

2.5.2. Induced Coupled Plasma Mass Spectroscopy (ICP-MS)

To achieve optimal results, the ICP-MS was operated under specific conditions, including a nebulizer gas flow of 0.84 L min$^{-1}$, plasma gas flow of 15 L min$^{-1}$, an auxiliary gas flow of 1.2 L min$^{-1}$, and a plasma radiofrequency power of 1600 W. The Syngistix 2.3 software was used to control the ICP-MS, and the tuning of all other parameters was carried out to optimize the Gd signal.

2.5.3. High-Performance Liquid Chromatography (HPLC)-ICP-MS

Flexar LC system (PerkinElmer) coupled with a Nexion 2000B (PerkinElmer) was used. The separation was executed using Superdex 75 increase 10/300 GL, and measurements were obtained through an isocratic mode of elution with acetate buffer 100 mM (pH 4.7) at a flow rate of 0.4 mL min$^{-1}$. The Gd signal was monitored utilizing isotopes 156 and 158 and Empower software version 7.3 was utilized to acquire the Gd signal.

### 2.5.4. Dynamic Light Scattering (DLS)

The hydrodynamic diameter distribution of the purified product was measured with a Zetasizer NanoS DLS instrument (laser He–Ne 633 nm) from Malvern Instruments.

2.5.5. Taylor Dispersion Analysis (TDA) TDA experiments were conducted using a TDA-ICP-MS hyphenation between a Sciex P/ACE MDQ instrument and a 7700 Agilent ICP-MS, described elsewhere.[53] Fused silica capillaries with an inner diameter of 75 µm and outer diameter of 375 µm, and a total length of 64 cm, were coated with hydroxypropylcellulose (HPC) using a solution of 0.05 g mL$^{-1}$ in water. Detection was carried out by ICP-MS at m/z=158 with a data acquisition rate of 500 ms point-1. Between runs, the capillary was flushed at 5 psi for 5 min with the mobilization medium. Peak deconvolution was carried out using Origin 8.5 software. The detailed method is described in ESI.

### 2.6. Competition ELISA

High-affinity 96-well plates were coated with either 2 µg mL$^{-1}$ PD-L1 or 1.25 µg mL-1 CD47 and allowed to adhere overnight at 4 °C. Plates were washed and blocked with a 10% FBS solution. Plates were incubated with sample VHH (A12 or A4) or NPs (AGuIX, AGuIX@A12, or AGuIX@A4) followed by incubation with 6.25 nM biotin-anti-PD-L1 or 50 nM biotin-anti-CD47. Plates were then incubated with avidin-HRP (Abcam ab7403, 1:40000 dilution) followed by TMB (ThermoFisher N301) to detect colorimetric changes. TMB conversation was stopped at 40 min and absorbance was read at 450 nm. The equilibrium inhibitory dissociation constant (Ki) was curve fitted using Graphpad 8.0.1 Top and Bottom are the plateaus in the units of the y-axis. LogK$_i$ is the log of the molar equilibrium dissociation constant of the unlabeled ligand (AGuIX, AGuIX@A12, or AGuIX@A4). RadioligandNM is the concentration of the labeled ligand (biotin-anti-PD-L1 or biotin-anti-CD47) and HotKdNM is the equilibrium dissociation constant of the labeled ligand (biotin-anti-PD-L1 or biotin-anti-CD47).

### 2.7. Biacore analyses

Biomolecular interactions between immobilized receptor PD-L1 and CD47 and analyte AGuIX@A12 or AGuIX@A4 were assessed by Surface Plasmon Resonance (SPR) on Biacore 2000 instrument (Cytiva).[53] For the kinetic assays, the A12 was injected up to a concentration of 61.2 nM, the A4 up to 67 nM, the AGuIX NPs at 0-915.7 nM in Gd$^{3+}$, the AGuIX@A12 VHH at 0-61.2 nM, (0-915.7 nM equivalent in Gd$^{3+}$) and the AGuIX@A4 VHH at 0-16.7 nM (0-240 nM equivalent in Gd$^{3+}$). The chip was prepared as described in SI. All characteristic interaction constants (equilibrium dissociation (KD) and kinetic rate of association and dissociation (kA and kD)) were determined by curve fitting using the Langmuir 1:1 binding model implemented in Biaevaluation software 4.1.1.[54] The detailed method is described in ESI.

### 2.8. Cell culture

Murine melanoma (wild-type B16F10 or hiPDL1-B16F10) cells were culture in 10% fetal bovine serum (FBS; Invitrogen, USA) and 1% pen/strep (10,000 U.mL-1 penicillin and 10,000 µg.mL$^{-1}$ streptomycin; Invitrogen, US) supplemented Roswell Park Memorial Institute medium (RPMI 1640; Gibco, Invitrogen, USA) at 37°C, 5% CO2 and optimal humidity.

### 2.9. Fluorescence Microscopy

Cover slips were placed in 48-well plates and seeded with either hiPDL1-B16 or wild-type B16 cells (50,000 cells/well) and allowed to attached overnight at 37°C. Cells were incubated with Cy5.5 (683/703 nm) tagged NPs (0.2 mg.mL$^{-1}$) for 1 h. Cells were fixed, blocked, and permeabilizing with a solution of 10% FBS and 0.3% Triton X-100. Cells were stained with CellMask Orange Plasma Membrane stain (ThermoFisher, 554/567 nm) and mounted with DAPI (350/470 nm) and Flouromount-G. Images were taken using a Zeiss AxioObserver microscope 63x. Fluorescence images were analyzed using ImageJ (version 2.14.0/1.54f) and corrected total cell fluorescence was calculated. [CTCF = Integrated Density$_{Cell}$ – (Area$_{Cell}$ x Mean fluorescence $_{Background}$)]

2.10. Animal tumor model

B16F10 cells (ATCC, USA) were cultured in DMEM media (Life Technologies, France) supplemented with 10% FBS (Life Technologies, France) and 1% penicillin/streptomycin (100 mg/mL), and 1% l-Glutamine, and maintained at 37 °C with 5% CO2 until confluence. Animal experiments have been performed according to the European directive 2010/63/EU and its transposition in the French law (Décret n° 2013-118). Experiments were conducted at the imaging facility CEA-SHFJ (authorization D91-471-105/ethics committee n°44). C57BL/6J mice (Janvier-Labs, France) were housed by 6 mice in each cage (bedding material: aspen wood) at room temperature 22°C, humidity 40%, under a regular 12-h dark/light cycle. Food and water were available ad libitum. 6 weeks old female C57BL/6 mice were purchased from Janvier laboratory. Mice were subcutaneously injected with 1 x 10$^6$ B16F10 cells suspended in DPBS (1× 100 µL) into both flanks while anesthetized with 2% isoflurane.

2.11. *Ex vivo* Biodistribution

The animals were anaesthetized with isoflurane (induction: 3%, maintenance: 1.5-2.0%) in a mixture of 100% O$_2$ (flow rate = 1.0- 1.5 L min$^{-1}$). At day 8 post-inoculation, 1.92 ± 0.09 MBq (mean ± SD), corresponding to the dose of 7.12 ± 0.35 and 2.29 ± 0.07 µg of AGuIX (mean ± SD) for AGuIX-[$^{89}$Zr]- and for AGuIX- [$^{89}$Zr]@A12 respectively[55], were intravenously injected (i.v.) into tumor-bearing mice (n = 8 for both AGuIX and AGuIX-A12, 22.0 ± 1.9 g per mouse). The mice were euthanized at 4 h and 24 h post-injection (n = 4 per time points) and tissue activity was determined for several organs of interest after harvesting (blood pool, intestines, kidneys, spleen, pancreas, liver, muscle, bone, brain, tumor). Activity in various organs of interest is represented in percentage of injected dose per gram (%ID/g) (Tab. S2).

2.12. Statistical analysis

The results are reported as mean ± SEM, as stated in the figure captions. Statistical analyses were conducted using GraphPad Prism (version 8.0.1). For comparisons involving three or more means, a one-way ANOVA followed by the Kruskal-Wallis nonparametric test (for non-Gaussian populations) was employed for in vitro internalization assays. All *in vitro* experiments were performed in triplicate. Mann-Whitney test for multiple comparisons was used to analyze ex vivo biodistribution data. Statistical significance was considered at p < 0.05.

3. **Results and discussion**

To evaluate the best approach for biofunctionalization of AGuIX NPs with VHH, we compared sortagging and click chemistry. The sortagging bioconjugation approach permits for the selection of specific sites for modification, enabling more accurate predictions regarding the biological impact of the chemical alteration. Modifying a protein terminus is expected to have a smaller effect on the protein's folding and functionality as well as the added advantage of defining the orientation of the immobilized protein, which is anticipated to better preserve its function compared to other immobilization techniques. This enzymatic reaction relies on the enzyme Sortase A (SrtA), a transpeptidase produced by Grampositive bacteria, which catalyzes the formation of a peptide bond between two peptides to attach specific proteins to the cell wall or pili assembly. [56–58] SrtA specifically recognizes the LPXTG sequence (with X representing any amino acid) and by nucleophilic attack makes it reactive to an N-terminal oligoglycine, forming a peptide bond between the threonine of LPXT and the oligoglycine. This natural ligation system has been repurposed for protein modification and has gained popularity as a research tool due to its high ligation selectivity, simplicity, robustness, and the availability of various SrtA variants (expressed in Escherichia coli or commercially available), as well as other required materials. [37,59] Sortagging reactions have been used to graft specific biomolecules, such as peptides or proteins, onto a protein or other biological targets of interest and has already proved efficacious in NP conjugation. [60–62] Click chemistry describes chemical reactions aimed at achieving high yield and high selectivity in the formation of carbon-heteroatom bond. Click chemistry reactions are based on the 1,3-dipolar cycloaddition of an azide and an alkyne to form a 1,2,3-triazole. This reaction has been widely utilized in various applications due to its simplicity in terms of preparation and purification steps, enabling rapid generation of new products with high reaction rates, such as bioconjugates.[60,63–65] The formation of the triazole linkage is irreversible and quantitative, providing excellent reaction stability. Azide and dibenzocyclooctyne (DBCO) are selected as strainpromoted azide-alkyne cycloaddition reactions (SPAAC) tools due to their relatively small functional groups with a favorable rate constant ($k_2$ = 0.2-0.5 $M^{-1}.s^{-1}$)[39] which have been shown to be effective. The crucial challenge of click chemistry is to preserve the integrity and functionality of the biomolecule after grafting. [39,66,67] Given the advantages and challenges associated with each method, we have synthesized AGuIX and VHH conjugates using both approaches to compare them.

### 3.1. Synthesis of AGuIX-VHH by sortagging

The sortagging reaction is based on enzymatic synthesis involving three primary steps: (1) introduction of maleimide (Mal) moieties on AGuIX NPs, (2) introduction of the peptide GGG(W/T)C using the Mal moieties, and (3) introduction of VHH through the transpeptidation enzymatic reaction (Fig. 1a).

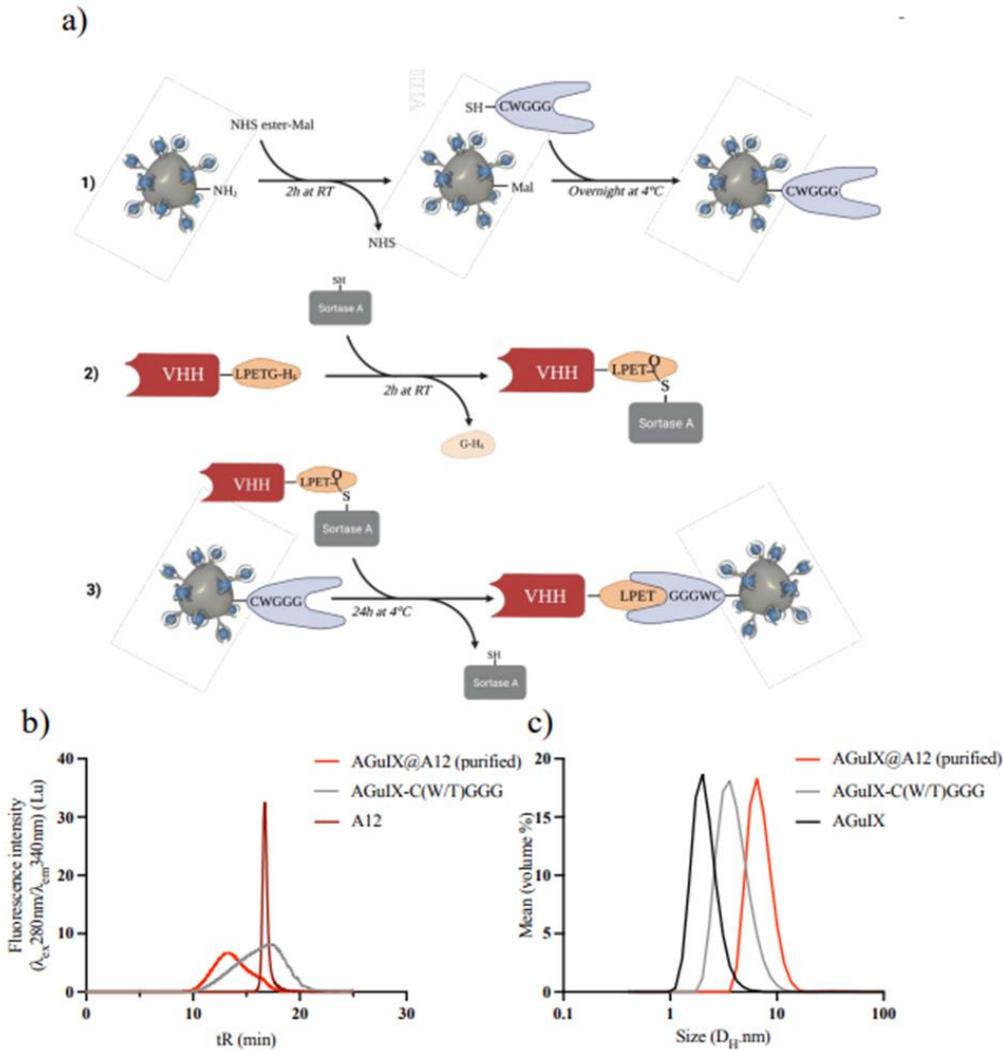

Fig. 1 Synthesis of AGuIX-VHH by sortagging reaction: method and characterizations (a) Scheme of the sortagging reaction, created with Biorender.com. (b) SEC chromatogram in the fluorescence intensity (λexc=280 nm; λem=340 nm) of the purified bioconjugate AGuIX@A12 (light red), AGuIX@GGG (grey) with A12 (dark red). (c) Measurement of hydrodynamic diameter by DLS of AGuIX@A12 (light red), AGuIX@GGG (grey) and AGuIX (black).

AGuIX NPs were modified to mimic the functionality of the Nterminal oligoglycine residue typically employed in sortagging. Initially, a Mal functional group was introduced onto the surface of AGuIX NPs using a bifunctional NHS/Mal linker. AGuIX-Mal was further modified with a short peptide linker that incorporates a cysteine amino acid at the C-terminal and a triglycine residue (GGG(W/T)C) at the N-terminal. To enable the use of sortagging, the VHH was engineered by incorporating a C-terminal StrA motif LPETGG and a 6xHis tag. The addition of the 6xHis tag motif facilitated protein purification and removal of the C-terminal residue released during sortagging.[20] The transpeptidation reaction between A12 and AGuIX-C(W/T)GGG was verified using SEC monitoring. After purification, the final product exhibited a retention time (tR) of 13.2 minutes, corresponding to a product size ranging from 29 to 44 kDa (Fig. 1b, Fig. S1.c). The equilibrium for the reaction was reached within 2-3 h at room temperature. Following isolation using Nibeads to capture and remove the 6xHis-containing reagents and by-products (StrA7m, residual peptides from VHH C-terminal, and unreacted VHH), AGuIX@A12 was purified via filtration and dispersed in PBS (Fig. S1.c). A VHH conversion of 32-33% was achieved during the sortagging process, resulting in an isolated yield of AGuIX@A12 of 5.3% (relative to A12) with a Gd/A12 ratio of 20 (Fig. 1b, Fig. S1.e). The hydrodynamic diameter (DH) of AGuIX@A12, as assessed by DLS, was 6.1 nm, slightly larger than that of AGuIX-C(W/T)GGG (4.1 nm) (Fig. 1c). This outcome confirms the modification of the particle surface and hints at the possibility of VHH grafting. After optimization, sortagging proved successful in grafting VHH onto the surface of AGuIX NPs. The key parameter to increase the grafting yield of VHH was to increase the equivalence of the oligoglycine in the reaction. However, this strategy also resulted in a higher NP/A12 ratio, which ideally should be around 1 for imaging applications. The NP/A12 ratio of 1 could be achieved after optimization of several parameters (Fig. S1.a to e), while VHH conversion decreased to 15%. The binding affinity, evaluated using the inhibitory constant (Ki) through a competitive enzyme-linked immunosorbent assay (ELISA), confirmed that the addition of AGuIX to A12 did not impact the competitive binding affinities (Fig. S2). Furthermore, AGuIX's lack of interaction with PD-L1 proteins also confirmed that the high affinity is solely attributed to the presence of the nanobody on AGuIX's surface. Moreover, the similar logKi values of 12.8 ± 1.1 nM, and 5.5 ± 0.6 nM for A12 and AGuIX@A12 obtained, respectively, strongly reinforced this result (Fig. S2).

### 3.2. Synthesis of AGuIX@VHH by click chemistry

Click chemistry based on strain-promoted azide-alkyne cycloaddition involves three primary steps: (1) introduction of the azide group on the VHH, (2) introduction of the DBCO group on the AGuIX NP, and (3) introduction of the VHH through the click chemistry reaction (Fig. 2.a). The first step of the synthesis involves functionalizing the VHH with the azide group based on NHS-ester reaction with lysine residues.[68–70] Functionalization was confirmed by SEC (Fig. S3.a) and matrix-assisted laser desorption/ionization-time of flight (MALDI-TOF) mass spectrometry (Fig. S3.b).[68]

The AGuIX-DBCO was prepared by the same reaction consisting in NHS reaction with the primary amines on the NP surface. After synthesis and purification, the modified NP presented a ratio of 10 Gd/DBCO, i.e. 1 to 2 DBCO functions per NP (Fig. S3.c and d). AGuIX-DBCO high excess ratio (450Gd/A12) was chosen for the click chemistry to minimize any reactant rate limitations and ensure complete reaction (Fig. 2b). At least 95% of the VHH in solution successfully reacted with AGuIX, resulting in the formation of AGuIX@A12 with a size ranging from approximately 29 to 44 kDa, as determined by Superdex 75 protein calibration (Fig. 2d; Fig. S3.e and f). Purification by preparative SEC was effective in removing the unreacted AGuIXDBCO and other by-products, isolating the AGuIX@A12 bioconjugate. Dynamic light scattering (DLS) and Taylor Dispersion Analysis (TDA) size measurements of the bioconjugate were consistent (Fig. 3c, Fig.S3.i). The analyses of hydrodynamic diameters using DLS and TDA exhibit coherence and complementarity. DLS offers a standard resolution, providing an average value, but it is limited in its ability to differentiate between species as such ultrasmall sizes. On the other hand, application of TDA to nanoparticles is a recent method that has proven to be effective in distinguishing various populations of ultrasmall size.[71]

An increase in hydrodynamic diameter was observed following biofunctionalization (in DLS, DH, AGuIX = 3.0 ± 1.1 nm vs DH, AGuIX@A12 = 5.4 ± 3.1 nm). TDA is a highly accurate and absolute method based on deconvolution methods, enabling the hydrodynamic diameters of AGuIX NPs to be determined via their diffusion coefficients.[53] This analysis revealed the presence of two populations within the AGuIX@A12 sample in which 75 ± 2% of AGuIX NPs were effectively functionalized with the VHH, while 25 ± 2% remained unfunctionalized. The minority population (25 ± 2%) are the remaining AGuIX-DBCO (Fig. S3.j). In summary, AGuIX@A12 was obtained with a 20Gd/A12 (approx. 1 NP/A12) and a reaction yield close to 20% (based on VHH). Just as with the sortagging product, the affinity of AGuIX@A12 was assessed using ELISA, resulting in a logK$_i$ value of 13.0 ± 0.4 nM, similar to A12.

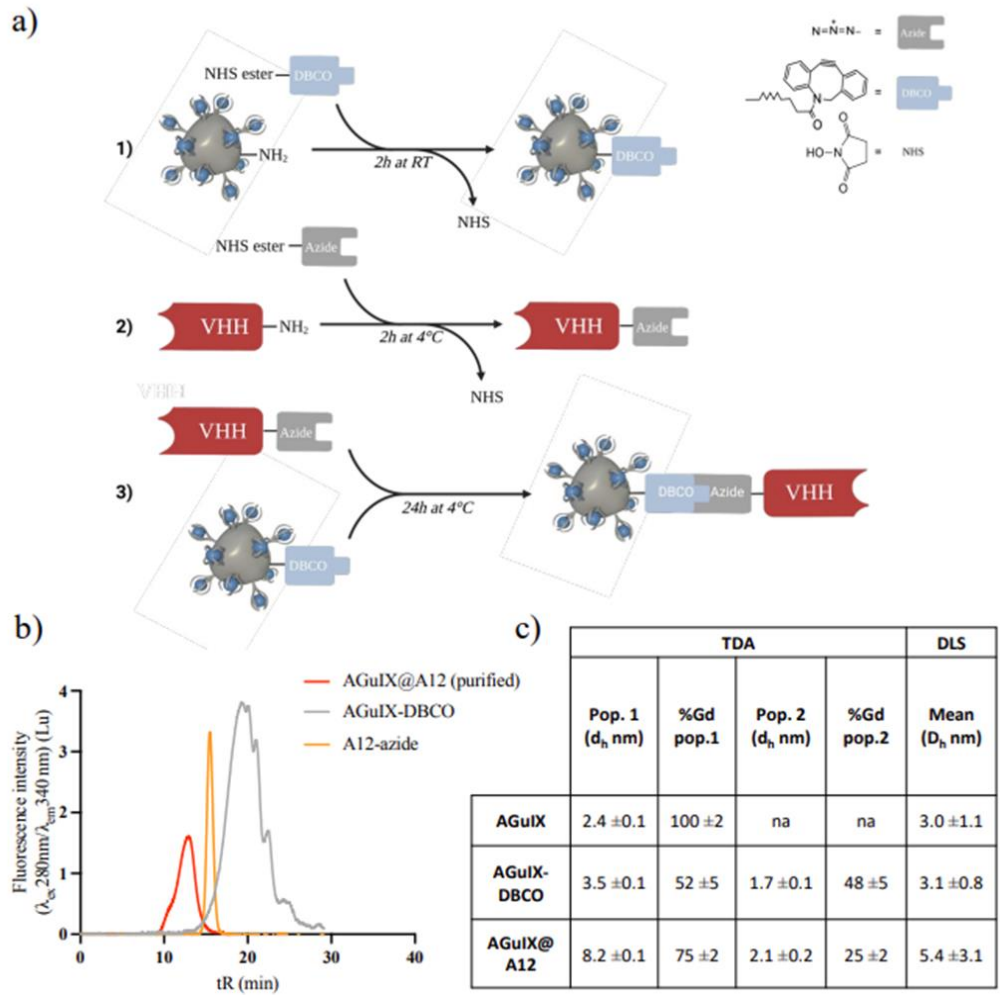

Fig. 2 Synthesis of AGuIX-VHH by click chemistry: method and characterizations (a) Scheme of the click chemistry reaction, created with Biorender.com. (b) SEC chromatogram in the fluorescence intensity (λexc=280 nm; λem=340 nm) of the purified bioconjugate AGuIX@A12 (red), AGuIX-DBCO (grey) and A12-azide (orange). (c) Comparison table of hydrodynamic diameters measured in TDA and DLS. Data presented as mean ± standard deviation.

The addition of AGuIX to the A12 nanobody through click chemistry did not influence competitive binding affinities; therefore, click chemistry did not compromise functional targeting properties (Fig. S2).

3.3. Methods comparison

First, we observed that both sortagging and click chemistry methods yielded AGuIX@VHH bioconjugates with relatively similar characteristics. The hydrodynamic size measurements ($D_{H,\,sortagging}$: 6.1 ± 3.7 nm and $D_{H,\,click}$: 5.4 ± 3.1 nm) and SEC chromatograms ($tR_{sortagging}$: 13.2 min and $tR_{click}$: 12.9 min) were consistent between the two methods. Furthermore, a crucial factor determining the reliability of the synthesis is the maintenance of the VHH's affinity with the PD-L1 after its grafting to AGuIX. Evaluation of this affinity through a competitive ELISA demonstrated that, in both synthesis approaches, the binding affinity remained robust and similar to the reference VHH ($logK_{isortagging}$: 5.5 ± 0.6 nM and $logK_{iclick}$: 13.0 ± 0.4 nM). Although both conjugation products maintained similar binding affinities for the PD-L1 ligand, the reactivity of VHH upon click reaction conditions was higher, with a grafting yield before purification $\eta_{click}$ > 95% (vs $\eta_{sortagging}$ < 50%). This observation aligns with existing literature, where sortagging reactivity efficacy may be limited due to equilibrium parameters (Fig S1.e).

[58,66,72] Moreover, the AGuIX-DBCO created during the click chemistry process could be separated from the final AGuIX@VHH bioconjugate, unlike the AGuIX-C(W/T)GGG used for sortagging. This separation provided a larger amount of purified AGuIX NP to be used for the click chemistry, reaching reaction equilibrium. Moreover, economic considerations are crucial when strategizing the scale-up of a process. It is noteworthy that biologics as used in the sortagging reaction are more expensive and less stable compared to chemical compounds. In conclusion, both sortagging and click chemistry methods afforded the AGuIX@VHH bioconjugates, however, the higher yield and lower cost associated with click chemistry have encouraged its adoption for further investigations (Table S1).

3.4. Proof of concept on A4 VHH

To demonstrate the reproducibility and robustness of the click chemistry approach, we conjugated a second VHH to AGuIX, this one specific to the CD47 receptor (A4. Mw =14.8 kDa). The first functionalization of the VHH with azide-PEG4-NHS demonstrated comparable efficacy to that of A12 (mean size of $MW_{A12-azide}$: 15.33 kDa (2 azide grafted) and $MW_{A4-azide}$:15.5 kDa (2-3 azide grafted)) (Fig. S6). The SEC indicate that the AGuIX@A4 (tR: 13 min) product has a size comparable to AGuIX@A12 (tR: 12.9 min) (Fig. 3).

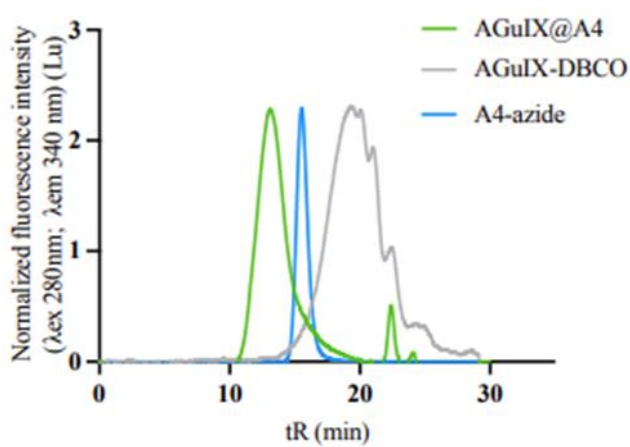

Fig. 3 Reproducibility of the bioconjugation by click chemistry protocol on the bioconjugation of AGuIX with A4 VHH. The SEC chromatogram in the fluorescence intensity (λexc = 280 nm; λem = 340 nm) shows the purified bioconjugates AGuIX@A4 (light green), AGuIX-DBCO (grey) and A4-azide (blue).

AGuIX@A4 was synthesized with a >95% conversion yield, similar to AGuIX@A12. After purification, the final purified product showed a ratio of 17Gd/VHH ratio, with a final yield of around 26%.

3.5. In vitro characterizations of AGuIX@VHH

A competitive ELISA confirmed that the addition of AGuIX to either VHH, A12 (Fig. 4a) or A4 (Fig. 4b), did not impact competition binding affinities. We demonstrated similar logKi values of 12.8 ± 1.1 nM and 13.0 ± 0.4 nM for A12 and AGuIX@A12, respectively, and 2.9 ± 0.2 nM and 3.3 ± 0.4 nM for A4 and AGuIX-A4, respectively. Moreover, AGuIX on its own exhibited no interaction with PD-L1 or CD47 ligands, providing additional support for the notion that the robust affinity observed primarily results from the incorporation of functional VHH on the AGuIX surface. These results were further validated using a second method: Biacore analysis which is based on a surface plasmon resonance (SPR) technique that which quantifies the association and dissociation phenomena. The affinity analyses conducted on the VHHs, and their associated receptors confirmed a strong binding affinity between these VHHs and their respective receptors, as shown by the equilibrium dissociation constants ($K_D$) (0.63 ± 0.01 nM for A12 on PD-L1 and 0.70 ± 0.13 nM for A4 on CD47) (Fig. 4c, Fig. S4-5). These results were supported by the literature, particularly in the case of A4, which has been extensively documented.[73] Comparing the $K_D$ values of AGuIX@VHH with their unmodified counterparts (0.39 ± 0.07 nM for AGuIX@A12 and 0.74 ± 0.01 nM for AGuIX@A4), we observed similar binding affinities further confirming that conjugation of AGuIX did not impact VHH binding. This similarity in $K_D$ values between AGuIX@VHH and unmodified VHH provided confirmation of the relevance of click chemistry in preserving the function of the VHHs. The A4, specific for the CD47 receptor, was selected to verify the specificity of the AGuIX@A12 to the PD-L1 receptor. The characteristic constants (equilibrium dissociation ($K_D$) and kinetic rate of association and dissociation ($k_A$ and $k_D$)) indicated a remarkable 10-fold stronger and faster binding affinity for A12 towards PD-L1 compared to A4, regardless of the bioconjugation status. These results confirm the specificity of the interaction between the nanobodies and their respective receptors (detailed in ESI). To confirm ligand-receptor interaction on cells, internalization and receptor blocking assays on melanoma B16F10 tumor cells, modified for high PD-L1 expression, were conducted to further validate the functionality of VHH bound to AGuIX (Fig. 4.d-e). Melanoma hiPDL1-B16F10 murine cells are highly metastatic, aggressive models mimicking patient phenotypes, expressing both PD-L1 and CD47.[51] The conjugation of A12 and A4 significantly increased internalization of AGuIX as seen in the representative images (Fig. 4d) and corrected total cell fluorescence (CTCF, Fig. 4e). Blocking with either anti-PD-L1 or anti-CD47 decreased internalization confirming specificity and functionality of the AGuIX@VHH. Studies were repeated in wildtype B16F10 cells (lower PD-L1 expression)[74](Fig. S7) with similar increase in internalization with the conjugation of A12 and A4 to AGuIX. Although complete loss of internalization was not observed, this may be an indication of only partial blocking of PD-L1 and CD47 receptors allowing for some receptor mediated internalization.

3.6. *Ex vivo* AGuIX@A12 targeting

Given the widespread use of PD-1 / PD-L1 inhibitors in clinical practice, we further evaluated the AGuIX@A12 product in proof-of-concept *ex vivo* studies. Targeting of PD-L1 was investigated in a murine melanoma model to confirm that AGuIX@A12 binding increasestumor accumulation. An *ex vivo* biodistribution study using zirconium-89 ($^{89}$Zr) radiolabeling was performed for accurate quantification of AGuIX and AGuIX@A12 in different organs at 4 h and 24 h post-intravenous injection (Fig. 4f - 4i). A short and long time point was selected based on previous preclinical biodistribution studies and current clinical trials involving AGuIX.

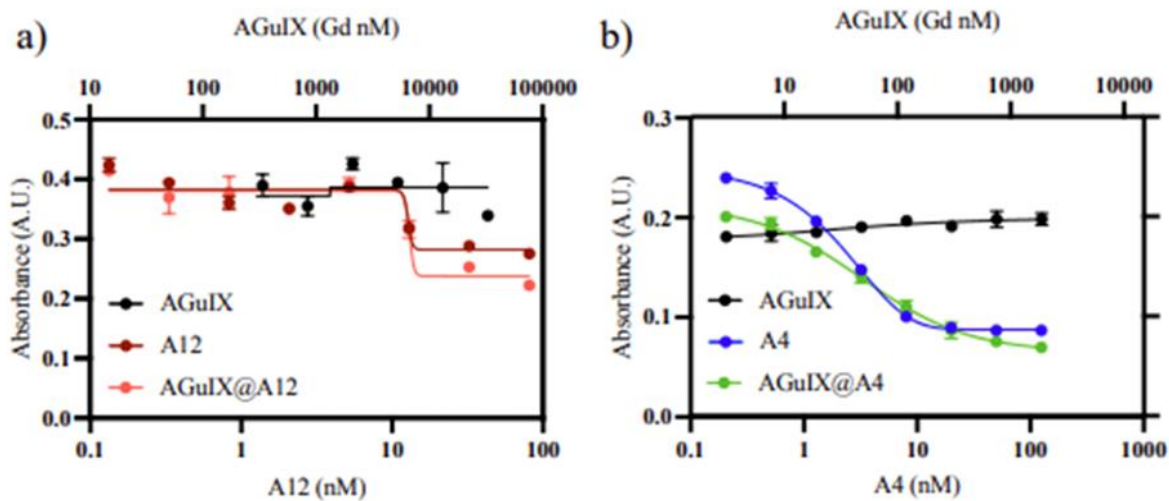

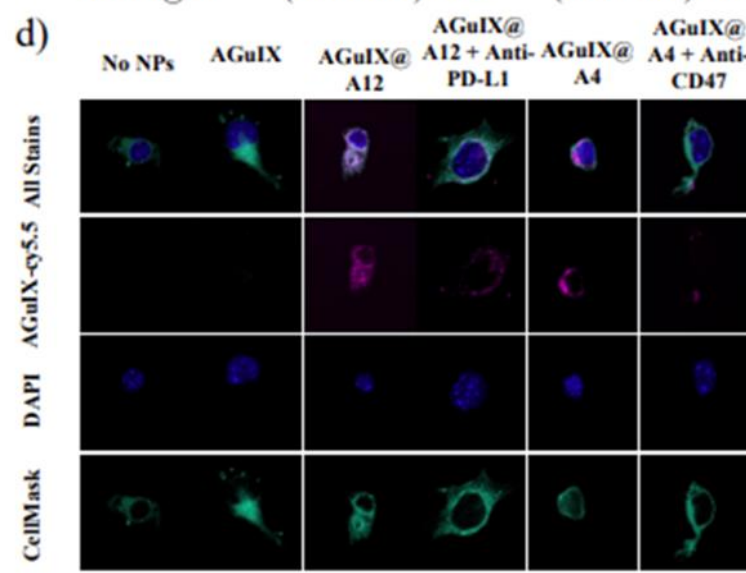
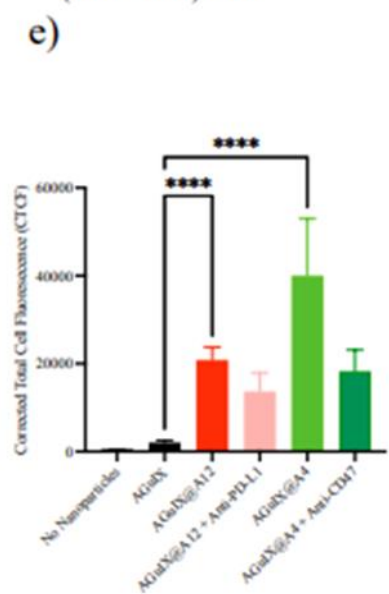

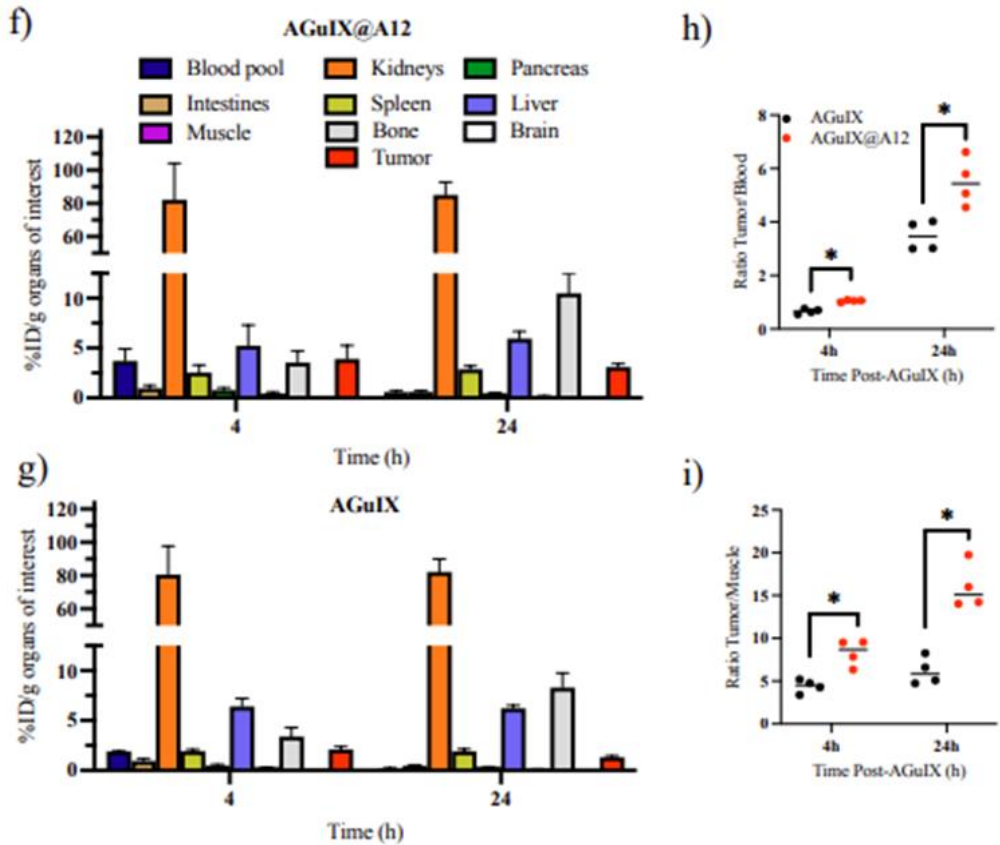

Fig. 4 Binding affinity of AGuIX@VHH conjugates. ELISA plates were either coated with (a) PD-L1 or (b) CD47 proteins and affinity of free and AGuIX@VHH bound VHH was assessed. No difference in binding affinity for A12 or A4 free VHH compared to AGuIX validating that functional targeting capabilities of VHH are not impacted by conjugation. Control of AGuIX further confirmed that binding is due to presence of VHH conjugates rather than non-specific NP and protein interactions (n = 3). (c) Table of binding affinity constants for AGuIX@VHH conjugates. All characteristic interaction constants ($K_D$, $k_A$ and $k_D$) between the A12 and A4 to their respectively PDL1 and CD47 receptors were determined by Langmuir 1:1 binding model. Triplicates were made for all measurements except A4@AGuIX only one measurement was made. (d) Internalization of AGuIX@VHH in high PD-L1 expressing melanoma cells. Metastatic hiPD-L1 B16F10 cells were incubated with either Cy5.5 conjugated AGuIX or AGuIX@VHH (magenta) for 1 h prior to staining with CellMask plasma membrane (green) and DAPI (blue) nuclear stain. For a subset of samples, PD-L1 and CD47 receptors were partially blocked with commercial anti-PD-L1 and anti-CD47, respectively, prior to AGuIX@VHH incubation. (e) AGuIX and AGuIX@VHH fluorescence was quantified using ImageJ to calculate corrected total cell fluorescence, CTCF (n = 11 – 40 cells, **** $p < 0.001$). (f,g) Quantitative analysis of *ex vivo* biodistribution (1.92 ± 0.09 MBq) at 4 h and 24 h after i.v. of (f) AGuIX@A12 and (g) AGuIX NPs which confirmed the specific active targeting of PD-L1 in B16F10 murine melanoma model. (n = 8; mean ± SD). (h) The tumor-to-blood and (i) the tumor-to-muscle ratios of AGuIX and AGuIX@A12 at different time points

Currently, in clinical trials, radiotherapy is delivered 4 hours after AGuIX administration, justifying the clinical relevance of a 4-hour biodistribution time point and multiple preclinical trials showed that tumor accumulation was increased up to 4h, while the 24-hour time point was chosen to study the expected longer persistence of targeted nanoparticles in the tumor.[1,5,75–77] As expected from previous work, both NPs accumulate within a few hours into the kidneys, leading to fast washout from the blood pool due to their ultrasmall size; lower accumulation of NPs in the liver can be detected.[76,78] Accumulation in the bone was attributed to the free [$^{89}$Zr] (non-coupled on the NPs)[79]. Both NPs can also accumulate in the tumor area due to the enhanced permeability and retention (EPR) effect, previously shown in several rodent models and in human studies.[6] However, a significantly higher accumulation of AGuIX-[$^{89}$Zr]@A12 was present in the tumor area compared to untargeted AGuIX-[$^{89}$Zr] (At 4 h, AGuIX: 2.10 ± 0.15 %ID.g$^{-1}$ vs AGuIX@A12: 3.90 ± 0.67 %ID.g$^{-1}$) (Fig. 4f - 4i). This difference was observed 4 h post-injection and was more pronounced 24 h postinjection (at 24h, AGuIX: 1.34 ± 0.06 %ID.g$^{-1}$ vs AGuIX@A12: 3.10 ± 0.15 %ID.g$^{-1}$). This higher accumulation and retention of AGuIX@VHH suggests targeting of the PD-L1 receptors at the surface of cancer cells, that is expected to increase active accumulation and slow the AGuIX-VHH washout within the tumor area. This *ex vivo* validation of the higher targeting of AGuIX@A12 vs AGuIX is evidence of the *in vivo* targeting of PD-L1 by AGuIX@A12. AGuIX@A12 is thus a promising candidate for future work involving MRI-detection of PD-L1 or for more targeted radiosensitization of PD-L1 expressing tumors.

## 4. Conclusion

We report the preparation of AGuIX@VHH bioconjugates using two distinct methods, sortagging and click chemistry. The two methods resulted in NPs with similar physicochemical characteristics using the A12 VHH. However, click chemistry exhibited higher VHH conversion while using lower-cost reagents, making it a more favourable approach for further investigations. Interestingly, the two bioconjugation methods yielded AGuIX@VHH with strong binding affinities, suggesting that site-specific VHH modification, such as sortagging, may not be required when coupling to relatively small NP such as AGuIX NP. PD-L1 targeting of AGuIX-A12 prepared by click chemistry was validated by *ex vivo* autoradiograpy demonstrating substantially greater uptake and persistence than untargeted NP. The A4 nanobody was used to the reliability and reproducibility of click synthesis for these constructs. Overall, click chemistry emerges as a favourable and promising approach for preparing a broad array of potential AGuIX@VHH bioconjugates that can be applied to multiple nanobodies, with diverse applications in targeted therapeutic interventions and imaging. Further comprehensive evaluation and validation in larger preclinical studies and clinical trials are necessary to fully explore the imaging and therapeutic potential of AGuIX@VHH and their role in enhancing cancer treatment.

**Supporting Information**

Supporting Information is available online or from the author.

**Author Contributions**

LC, GB and PR designed and performed the chemical experiments and product characterization. LS, ZM and NB conducted in vitro ELISA experiments and microscopy imaging. LH performed the Biacore experiments. AH conducted the TDA-ICP-MS experiments. VG-C carried out SEC preparative purifications. FD conducted the MALDITOF experiments. CT performed the ex vivo experiments. TD helped with NP formulation and characterization. LC, GB and NB wrote the manuscript. FL, OT, MD, JS, GB, NB and RB conceptualized the idea, supervised the study, and revised the manuscript. All authors discussed the results and approved the final version of the manuscript.

Conflicts of interest

LC is an employee of NH TherAGuIX that develops AGuIX NPs. OT and FL possess shares in this company. JS reports research support paid to the institution: Merck, BMS, Regeneron, Debiopharm, EMD, Seronoand consulting / Scientific Advisory Board / Travel fees: Castle Biosciences, Genentech, Immunitas, Debiopharm, BMS, LEK, Catenion, ACI Clinical, Astellas, Stimit, Merck KGA, SIRPant, EMD Serono. Stock options: Immunitas.

Acknowledgements Special thanks to the international IRP Harvard-Lyon Radioboost project, which aims to boost radiotherapy with ultra-small nanoparticles.

Support for this project was provided by Viewray, Inc (agreement #2019A017789).

**References**

1 C. Verry, S. Dufort, J. Villa, M. Gavard, C. Iriart, S. Grand, J. Charles, B. Chovelon, J.-L. Cracowski, J.-L. Quesada, C. Mendoza, L. Sancey, A. Lehmann, F. Jover, J.-Y. Giraud, F. Lux, Y. Crémillieux, S. McMahon, P. J. Pauwels, D. Cagney, R. Berbeco, A. Aizer, E. Deutsch, M. Loeffler, G. Le Duc, O. Tillement and J. Balosso, Radiother. Oncol., 2021, 160, 159–165.

2 E. Thivat, M. Casile, J. Moreau, I. Molnar, S. Dufort, K. Seddik, G. Le Duc, O. De Beaumont, M. Loeffler, X. Durando and J. Biau, BMC Cancer, 2023, 23, 344.

3 F. Lux, V. L. Tran, E. Thomas, S. Dufort, F. Rossetti, M. Martini, C. Truillet, T. Doussineau, G. Bort, F. Denat, F. Boschetti, G. Angelovski, A. Detappe, Y. Crémillieux, N. Mignet, B.-T. Doan, B. Larrat, S. Meriaux, E. Barbier, S. Roux, P. Fries, A. Müller, M.-C. Abadjian, C. Anderson, E. Canet-Soulas, P. Bouziotis, M. Barberi-Heyob, C. Frochot, C. Verry, J. Balosso, M. Evans, J. Sidi-Boumedine, M. Janier, K. Butterworth, S. McMahon, K. Prise, M.-T. Aloy, D. Ardail, C. Rodriguez-Lafrasse, E. Porcel, S. Lacombe, R. Berbeco, A. Allouch, J.-L. Perfettini, C. Chargari, E. Deutsch, G. Le Duc and O. Tillement, Br. J. Radiol., 2019, 92, 20180365.

4 F. Lux, L. Sancey, A. Bianchi, Y. Crémillieux, S. Roux and O. Tillement, Nanomed., 2015, 10, 1801–1815.

5 L. Carmès, M. Banerjee, P. Coliat, S. Harlepp, X. Pivot, O. Tillement, F. Lux and A. Detappe, Adv. Ther., 2023, n/a, 2300019.

6 G. Bort, F. Lux, S. Dufort, Y. Crémillieux, C. Verry and O. Tillement, Theranostics, 2020, 10, 1319–1331.

7 A. Detappe, C. Mathieu, C. Jin, M. P. Agius, M.-C. Diringer, V.-L. Tran, X. Pivot, F. Lux, O. Tillement, D. Kufe and P. P. Ghoroghchian Int. J. Radiat. Oncol., 2020, 108, 1380–1389.

8 A. Detappe, M. Reidy, Y. Yu, C. Mathieu, H. V.-T. Nguyen, T. P. Coroller, F. Lam, P. Jarolim, P. Harvey, A. Protti, Q.-D. Nguyen, J. A. Johnson, Y. Cremillieux, O. Tillement, I. M. Ghobrial and P. P. Ghoroghchian, Nanoscale, 2019, 11, 20485–20496.

9 E. Thomas, C. Mathieu, P. Moreno-Gaona, V. Mittelheisser, F. Lux, O. Tillement, X. Pivot, P. P. Ghoroghchian and A. Detappe, Adv. Healthc. Mater., 2022, 11, 2101565.

10 M. M. Gomari, S. Abkhiz, T. G. Pour, E. Lotfi, N. Rostami, F. N. Monfared, B. Ghobari, M. Mosavi, B. Alipour and N. V. Dokholyan, Mol. Med., 2022, 28, 146.

11 P. Chames, M. Van Regenmortel, E. Weiss and D. Baty, Br. J. Pharmacol., 2009, 157, 220–233.

12 S. Sun, Z. Ding, X. Yang, X. Zhao, M. Zhao, L. Gao, Q. Chen, S. Xie, A. Liu, S. Yin, Z. Xu and X. Lu, Int. J. Nanomedicine, 2021, 16, 2337–2356.

13 A. Muruganandam, J. Tanha, S. Narang and D. Stanimirovic, FASEB J. Off. Publ. Fed. Am. Soc. Exp. Biol., 2002, 16, 240–242.

14 J. Wesolowski, V. Alzogaray, J. Reyelt, M. Unger, K. Juarez, M. Urrutia, A. Cauerhff, W. Danquah, B. Rissiek, F. Scheuplein, N. Schwarz, S. Adriouch, O. Boyer, M. Seman, A. Licea, D. V. Serreze, F. A. Goldbaum, F. Haag and F. Koch-Nolte, Med. Microbiol. Immunol. (Berl.), 2009, 198, 157–174.

15 M. Liu, Y. Zhu, T. Wu, J. Cheng and Y. Liu, Chem. – Eur. J., 2020, 26, 7442–7450.

16 E. Y. Yang and K. Shah, Front. Oncol., 2020, 10, 1182.

17 T. De Meyer, S. Muyldermans and A. Depicker, Trends Biotechnol., 2014, 32, 263–270.

18 P. Bannas, J. Hambach and F. Koch-Nolte, Front. Immunol.

19 M. Dougan, J. R. Ingram, H.-J. Jeong, M. M. Mosaheb, P. T. Bruck, L. Ali, N. Pishesha, O. Blomberg, P. M. Tyler, M. M. Servos, M. Rashidian, Q.-D. Nguyen, U. H. von Andrian, H. L. Ploegh and S. K. Dougan, Cancer Immunol. Res., 2018, 6, 389–401.

20 J. R. Ingram, O. S. Blomberg, J. T. Sockolosky, L. Ali, F. I. Schmidt, N. Pishesha, C. Espinosa, S. K. Dougan, K. C. Garcia, H. L. Ploegh and M. Dougan, Proc. Natl. Acad. Sci., 2017, 114, 10184–10189.

21 J. R. Ingram, M. Dougan, M. Rashidian, M. Knoll, E. J. Keliher, S. Garrett, S. Garforth, O. S. Blomberg, C. Espinosa, A. Bhan, S. C. Almo, R. Weissleder, H. Lodish, S. K. Dougan and H. L. Ploegh, Nat. Commun., 2017, 8, 647.

22 L. Fehrenbacher, A. Spira, M. Ballinger, M. Kowanetz, J. Vansteenkiste, J. Mazieres, K. Park, D. Smith, A. Artal-Cortes, C. Lewanski, F. Braiteh, D. Waterkamp, P. He, W. Zou, D. S. Chen, J. Yi, A. Sandler and A. Rittmeyer, The Lancet, 2016, 387, 1837–1846.

23 D. R. Spigel, C. Faivre-Finn, J. E. Gray, D. Vicente, D. Planchard, L. Paz-Ares, J. F. Vansteenkiste, M. C. Garassino, R. Hui, X. Quantin, A. Rimner, Y.-L. Wu, M. Özgüroğlu, K. H. Lee, T. Kato, M. de Wit, T. Kurata, M. Reck, B. C. Cho, S. Senan, J. Naidoo, H. Mann, M. Newton, P. Thiyagarajah and S. J. Antonia, J. Clin. Oncol., 2022, 40, 1301–1311.

24 A. Bouleau, V. Lebon and C. Truillet, Pharmacol. Ther., 2021, 222, 107786.

25 Y.-C. Chen, W. Shi, J.-J. Shi and J.-J. Lu, J. Cancer Res. Clin. Oncol., 2022, 148, 1–14.

26 E. C. Piccione, S. Juarez, J. Liu, S. Tseng, C. E. Ryan, C. Narayanan, L. Wang, K. Weiskopf and R. Majeti, mAbs, 2015, 7, 946–956.

27 M. P. Chao, C. H. Takimoto, D. D. Feng, K. McKenna, P. Gip, J. Liu, J.-P. Volkmer, I. L. Weissman and R. Majeti, Front. Oncol., 2019, 9, 1380.

28 H. A. Burris III, A. I. Spira, M. H. Taylor, O. O. Yeku, J. F. Liu, P. N. Munster, E. P. Hamilton, J. S. Thomas, F. Gatlin, R. T. Penson, T. A. Abrams, M. S. Dhawan, J. M. Walling, J. W. Frye, K. Romanko, V. Sung, C. Brachmann and A. B. El-Khoueiry, J. Clin. Oncol., 2021, 39, 2516– 2516.

29 Z. Jiang, H. Sun, J. Yu, W. Tian and Y. Song, J. Hematol. Oncol.J Hematol Oncol, 2021, 14, 180.

30 I. E. Krop, N. Masuda, T. Mukohara, S. Takahashi, T. Nakayama, K. Inoue, H. Iwata, T. Toyama, Y. Yamamoto, D. M. Hansra, M. Takahashi, A. Osaki, K. Koyama, T. Inoue, T. Yonekura, J. Mostillo, S. Ohwada, Y. Tanaka, D. W. Sternberg and K. Yonemori, J. Clin. Oncol., 2022, 40, 1002–1002.

31 R. K. Vaddepally, P. Kharel, R. Pandey, R. Garje and A. B. Chandra, Cancers, 2020, 12, 738.

32 H. Sato, N. Okonogi and T. Nakano, Int. J. Clin. Oncol., 2020, 25, 801–809.

33 L. Wen, F. Tong, R. Zhang, L. Chen, Y. Huang and X. Dong, Front. Oncol.

34 Y. Nishiga, A. P. Drainas, M. Baron, D. Bhattacharya, A. A. Barkal, Y. Ahrari, R. Mancusi, J. B. Ross, N. Takahashi, A. Thomas, M. Diehn, I. L. Weissman, E. E. Graves and J. Sage, Nat. Cancer, 2022, 3, 1351– 1366.

35 E. Rostami, M. Bakhshandeh, H. Ghaffari-Nazari, M. Alinezhad, M. Alimohammadi, R. Alimohammadi, G. Mahmoodi Chalbatani, E. Hejazi, T. J. Webster, J. Tavakkol-Afshari and S. A. Jalali, PLoS ONE, 2022, 17, e0273547.

36 Y. Zhang, K.-Y. Park, K. F. Suazo and M. D. Distefano, Chem. Soc. Rev., 2018, 47, 9106–9136.

37 X. Dai, A. Böker and U. Glebe, RSC Adv., 2019, 9, 4700–4721.

38 M. Ritzefeld, Chem. – Eur. J., 2014, 20, 8516–8529.

39 L. Taiariol, C. Chaix, C. Farre and E. Moreau, Chem. Rev., 2021, acs.chemrev.1c00484.

40 J. M. Eeftens, J. van der Torre, D. R. Burnham and C. Dekker, BMC Biophys., 2015, 8, 9.

41 T. I. Chio and S. L. Bane, Methods Mol. Biol. Clifton NJ, 2020, 2078, 83–97.

42 J. M. Baskin, J. A. Prescher, S. T. Laughlin, N. J. Agard, P. V. Chang, I. A. Miller, A. Lo, J. A. Codelli and C. R. Bertozzi, Proc. Natl. Acad. Sci., 2007, 104, 16793–16797.

43 E. Kim and H. Koo, Chem. Sci., 2019, 10, 7835–7851.

44 B. Stump, ChemBioChem, 2022, 23, e202200016.

45 Explaining Nobel Prize-winning chemistry techniques, https://www.buffalo.edu/ubnow/stories/2022/10/nobel-prizechemistry.html, (accessed March 9, 2023).

46 G. Le Duc, S. Roux, A. Paruta-Tuarez, S. Dufort, E. Brauer, A. Marais, C. Truillet, L. Sancey, P. Perriat, F. Lux and O. Tillement, Cancer Nanotechnol., 2014, 5, 4.

47 A. Mignot, C. Truillet, F. Lux, L. Sancey, C. Louis, F. Denat, F. Boschetti, L. Bocher, A. Gloter, O. Stéphan, R. Antoine, P. Dugourd, D. Luneau, G. Novitchi, L. C. Figueiredo, P. C. de Morais, L. Bonneviot, B. Albela, F. Ribot, L. Van Lokeren, I. Déchamps-Olivier, F. Chuburu, G. Lemercier, C. Villiers, P. N. Marche, G. Le Duc, S. Roux, O. Tillement and P. Perriat, Chem. – Eur. J., 2013, 19, 6122–6136.

48 N. Brown, P. Rocchi, L. Carmès, R. Guthier, M. Iyer, L. Seban, T. Morris, S. Bennett, M. Lavelle, J. Penailillo, R. Carrasco, C. Williams, E. Huynh, Z. Han, E. Kaza, T. Doussineau, S. M. Toprani, X. Qin, Z. D. Nagel, K. A. Sarosiek, A. Hagège, S. Dufort, G. Bort, F. Lux, O. Tillement and R. Berbeco, Theranostics, 2023, 13, 4711–4729.

49 E. Thomas, L. Colombeau, M. Gries, T. Peterlini, C. Mathieu, N. Thomas, C. Boura, C. Frochot, R. Vanderesse, F. Lux, M. BarberiHeyob and O. Tillement, Int. J. Nanomedicine, 2017, Volume 12, 7075–7088.

50 C. Truillet, E. Thomas, F. Lux, L. T. Huynh, O. Tillement and M. J. Evans, Mol. Pharm., 2016, 13, 2596–2601.

51 E. Thomas, C. Mathieu, P. Moreno-Gaona, V. Mittelheisser, F. Lux, O. Tillement, X. Pivot, P. P. Ghoroghchian and A. Detappe, Adv. Healthc. Mater., 2022, 11, 2101565.

52 E. Thomas, Cancer, Université de Lyon, 2017.

53 L. Labied, P. Rocchi, T. Doussineau, J. Randon, O. Tillement, H. Cottet, F. Lux and A. Hagège, Anal. Chim. Acta, 2021, 1185, 339081.

54 L. Heinrich, N. Tissot, D. J. Hartmann and R. Cohen, J. Immunol. Methods, 2010, 352, 13–22.

55 V.-L. Tran, F. Lux, N. Tournier, B. Jego, X. Maître, M. Anisorac, C. Comtat, S. Jan, K. Selmeczi, M. J. Evans, O. Tillement, B. Kuhnast and C. Truillet, Adv. Healthc. Mater., 2021, 10, e2100656.

56 T. Spirig, E. M. Weiner and R. T. Clubb, Mol. Microbiol., 2011, 82, 1044–1059.

57 H. E. Morgan, W. B. Turnbull and M. E. Webb, Chem. Soc. Rev., 2022, 51, 4121–4145.

58 J. M. Antos, M. C. Truttmann and H. L. Ploegh, Curr. Opin. Struct. Biol., 2016, 38, 111–118.

59 J. E. Glasgow, M. L. Salit and J. R. Cochran, J. Am. Chem. Soc., 2016, 138, 7496–7499.

60 S. A. M. van Lith, S. M. J. van Duijnhoven, A. C. Navis, W. P. J. Leenders, E. Dolk, J. W. H. Wennink, C. F. van Nostrum and J. C. M. van Hest, Bioconjug. Chem., 2017, 28, 539–548.

61 L. Liu, J. L. Gray, E. W. Tate and A. Yang, Trends Biotechnol., , DOI:10.1016/j.tibtech.2023.05.004.

62 M. W. Popp, J. M. Antos, G. M. Grotenbreg, E. Spooner and H. L. Ploegh, Nat. Chem. Biol., 2007, 3, 707–708.

63 N. Kotagiri, Z. Li, X. Xu, S. Mondal, A. Nehorai and S. Achilefu, Bioconjug. Chem., 2014, 25, 1272–1281.

64 V. Aragon-Sanabria, A. Aditya, L. Zhang, F. Chen, B. Yoo, T. Cao, B. Madajewski, R. Lee, M. Z. Turker, K. Ma, S. Monette, P. Chen, J. Wu, S. Ruan, M. Overholtzer, P. Zanzonico, C. M. Rudin, C. Brennan, U. Wiesner and M. S. Bradbury, Clin. Cancer Res., 2022, 28, 2938– 2952.

65 L. Williams, L. Li, P. J. Yazaki, P. Wong, A. Miller, T. Hong, E. K. Poku, S. Bhattacharya, J. E. Shively and M. Kujawski, Biotechnol. J., 2023, e2300115.

66 A. Debon, E. Siirola and R. Snajdrova, JACS Au, 2023, 3, 1267– 1283.

67 C. J. Pickens, S. N. Johnson, M. M. Pressnall, M. A. Leon and C. J. Berkland, Bioconjug. Chem., 2018, 29, 686– 701.

68 V. Solntceva, M. Kostrzewa and G. Larrouy-Maumus, Front. Cell. Infect. Microbiol., 2021, 10, 621452.

69 S. Mädler, C. Bich, D. Touboul and R. Zenobi, J. Mass Spectrom., 2009, 44, 694–706.

70 J. S. Nanda and J. R. Lorsch, in Methods in Enzymology, ed. J. Lorsch, Academic Press, 2014, vol. 536, pp. 87–94.

71 L. Labied, P. Rocchi, T. Doussineau, J. Randon, O. Tillement, H. Cottet, F. Lux and A. Hagège, Anal. Chim. Acta, 2021, 1185, 339081.

72 Q. Wu, H. L. Ploegh and M. C. Truttmann, ACS Chem. Biol., 2017, 12, 664–673.

73 J. T. Sockolosky, M. Dougan, J. R. Ingram, C. C. M. Ho, M. J. Kauke, S. C. Almo, H. L. Ploegh and K. C. Garcia, Proc. Natl. Acad. Sci. U. S. A., 2016, 113, E2646–E2654.

74 A. Quijano-Rubio, A. M. Bhuiyan, H. Yang, I. Leung, E. Bello, L. R. Ali, K. Zhangxu, J. Perkins, J.-H. Chun, W. Wang, M. J. Lajoie, R. Ravichandran, Y.-H. Kuo, S. K. Dougan, S. R. Riddell, J. B. Spangler, M. Dougan, D.-A. Silva and D. Baker, Nat. Biotechnol., 2023, 41, 532– 540.

75 S. Kotb, A. Detappe, F. Lux, F. Appaix, E. L. Barbier, V.-L. Tran, M. Plissonneau, H. Gehan, F. Lefranc, C. Rodriguez-Lafrasse, C. Verry, R. Berbeco, O. Tillement and L. Sancey, Theranostics, 2016, 6, 418–427.

76 E. Thivat, M. Casile, J. Moreau, I. Molnar, S. Dufort, K. Seddik, G. Le Duc, O. De Beaumont, M. Loeffler, X. Durando and J. Biau, BMC Cancer, 2023, 23, 344.

77 C. Verry, L. Sancey, S. Dufort, G. Le Duc, C. Mendoza, F. Lux, S. Grand, J. Arnaud, J. L. Quesada, J. Villa, O. Tillement and J. Balosso, BMJ Open, 2019, 9, e023591.

78 F. Lux, V. L. Tran, E. Thomas, S. Dufort, F. Rossetti, M. Martini, C. Truillet, T. Doussineau, G. Bort, F. Denat, F. Boschetti, G. Angelovski, A. Detappe, Y. Crémillieux, N. Mignet, B.-T. Doan, B. Larrat, S. Meriaux, E. Barbier, S. Roux, P. Fries, A. Müller, M.-C. Abadjian, C. Anderson, E. Canet-Soulas, P. Bouziotis, M. Barberi-Heyob, C. Frochot, C. Verry, J. Balosso, M. Evans, J. Sidi-Boumedine, M. Janier, K. Butterworth, S. McMahon, K. Prise, M.-T. Aloy, D. Ardail, C. Rodriguez-Lafrasse, E. Porcel, S. Lacombe, R. Berbeco, A. Allouch, J.- L. Perfettini, C. Chargari, E. Deutsch, G. Le Duc and O. Tillement, Br. J. Radiol., 2019, 92, 20180365.

79 D. S. Abou, T. Ku and P. M. Smith-Jones, Nucl. Med. Biol., 2011, 38, 675–681.